\documentclass[twocolumn,10pt,prd,aps,showpacs]{revtex4}
\def \D {\mbox{D}}
\def \D {\mbox{D}}

\def \curl {\mbox{curl}\,}

\def\3nab{\tilde{\nabla}}

\def\hs {\,-\,}

\def\be {\begin{equation}}
\def\ee {\end{equation}}
\def\bea {\begin{eqnarray}}
\def\eea {\end{eqnarray}}

\newcommand{\sfrac}[2]{{\textstyle{#1\over#2}}}
\begin{document}
\title{Large\hs scale perturbations on the brane \\ and the isotropy 
of the cosmological singularity}
\author{Naureen Goheer${}^1$, Peter K. S. Dunsby${}^{1,2}$, Alan
Coley${}^{3}$ and Marco Bruni${}^{4}$}
\affiliation{${}^1$ Department of Mathematics and Applied Mathematics,
University of Cape Town, Rondebosch 7701, Cape Town, South Africa.}
\affiliation{${}^2$ South African Astronomical Observatory, Observatory
7925, Cape Town, South Africa.}
\affiliation{${}^3$ Department of Mathematics and Statistics,
Dalhousie University, Halifax, Nova Scotia, Canada B3H 3JS.}
\affiliation{${}^4$ Institute of Cosmology and Gravitation, University of
Portsmouth, Mercantile House, Portsmouth PO1 2EG.}
\date{July 22, 2004}
\begin{abstract}
We present the complete set of propagation and constraint equations for the
kinematic and non\hs local first order quantities which describe general 
linear inhomogeneous and anisotropic perturbations of a flat FRW braneworld 
with vanishing cosmological constant and decompose them in the standard way 
into their scalar, vector and tensor contributions. A detailed analysis of 
the perturbation dynamics is performed using dimensionless variables that 
are specially tailored for the different regimes of interest; namely, 
the {\it low energy GR regime}, the {\it high energy regime} and 
the {\it dark energy regime}. Tables are presented for the evolution of 
all the physical quantities, making it easy to do a detailed comparison 
of the past asymptotic behaviour of the perturbations of these models. 
We find results that exactly match those obtained in the analysis of the 
spatially inhomogeneous $G_{2}$ braneworld cosmologies presented recently; 
i.e., that isotropization towards the ${\cal F}_b$ model occurs for 
$\gamma > 4/3$.
\end{abstract}
\maketitle
\section{Introduction}
A well known problem of cosmology is to explain the very high
degree of isotropy observed in the Cosmic Microwave Background (CMB). 
In a theory such as general relativity, where isotropy is a special 
rather than generic feature of cosmological models, we need a dynamical
mechanism able to produce isotropy. Inflation was proposed, among
other reasons, as a way to isotropize the universe. Inflation is
effective in this sense, but it needs homogeneous enough initial
data in order for inflation to begin \cite{KT}. Although one could perhaps 
adopt the view that one smooth enough patch in an otherwise non\hs smooth 
initial universe is all that is needed to explain observations, 
this may not be satisfactory \cite{linde}: the isotropy problem 
remains open in standard cosmology.

Recently, a number of authors
\cite{MS,SC2,SVF,SC1,coley2,coley1,HCY,HI,naureen1,naureen2}
have addressed the issue of isotropization in the context of braneworld
cosmology based on a generalization of the Randall and Sundrum
model \cite{RS,SMS}. Here the bulk is 5\hs dimensional and contains
only a cosmological constant, assumed to be negative (see \cite{roy2} for
a comprehensive review).

In all cases considered, an interesting result was found: unlike
general relativity, where in general the cosmological singularity
is anisotropic, the past attractor for spatially homogeneous 
anisotropic models in the brane is a simple Robertson\hs Walker (RW) 
model ${\cal F}_b$ \cite{coley2,langlois}. Since this result was also 
found to hold for Bianchi IX models \cite{coley2,coley1} as well as for some 
inhomogeneous models, the author suggested that the isotropic 
singularity could be a generic feature of brane cosmological models.

In a recent paper \cite{CHL}, this conjecture was supported by studying
the dynamics of a class of {\em spatially inhomogeneous $G_{2}$}
cosmological models in the braneworld scenario. A {\it numerical}
analysis of the governing system of evolution equations led to the
result that for $\gamma > 4/3$ isotropization towards a simple RW model
${\cal F}_b$ occurs as $\tau\rightarrow -\infty$ for {\em all} 
initial conditions. In the case of radiation ($\gamma=4/3$), the models 
were still found to isotropize as $\tau\rightarrow -\infty$, albeit 
slowly. It can therefore be concluded that an initial isotropic singularity 
occurs in all of these $G_2$ spatially inhomogeneous brane cosmologies for 
a range of parameter values which include the physically important cases 
of radiation and a scalar field source. The numerical results were 
confirmed by a qualitative dynamical analysis and a detailed calculation 
of the past asymptotic decay rates \cite{CHL}.

A similar result is also obtained in a related perturbative study where a
careful analysis of generic linear inhomogeneous and anisotropic
perturbations of the ${\cal F}_b$ model \cite{dunsbyetal} was conducted.
Solutions were obtained for the large\hs scale evolution of {\it scalar},
{\it vector} and {\it tensor} perturbations showing that the ${\cal F}_b$ model is
stable in the past (as $\tau\rightarrow -\infty$) with respect to 
generic inhomogeneous and anisotropic perturbations provided the matter 
is described by a non\hs inflationary perfect fluid with $\gamma$\hs law 
equation of state parameter satisfying $\gamma > 1$. In particular, it 
was shown that the expansion normalised shear vanishes 
as $\tau\rightarrow -\infty$, signalling isotropization.

Brane cosmology thus has the very attractive feature of having isotropy
built in, and although inflation in this context would still be the most likely
way of producing the fluctuations seen in the CMB, there would be no need for
special initial conditions for it to start. Also, the Penrose
conjecture \cite{penrose} on gravitational entropy and an initially vanishing
measure of the Weyl tensor might be satisfied, c.f. \cite{tod}.

The aim of this paper is to give a more compressive large\hs scale 
perturbative analysis of flat Friedmann\hs Robertson\hs  Walker (FRW) 
brane models with vanishing cosmological constant by combining the 
{\it high energy} results in \cite{dunsbyetal} with an analysis of the 
other important stages in the braneworld evolution, namely the 
{\it low energy GR} and {\it dark energy} regimes. To make this 
precise we define dimensionless variables that are specially tailored 
for each regime of interest. In this way we are able to clarify further 
the past asymptotic behaviour of these models and obtain results which 
match the analysis of the spatially inhomogeneous $G_{2}$ cosmologies 
presented in \cite{CHL}; i.e., that isotropization towards the ${\cal F}_b$ 
model occurs for $\gamma > 4/3$.

The paper is organised as follows. In section II we will give a
brief summary of the braneworld scenario and the induced field equations
on the brane. In section III we introduce dimensionless expansion normalised
variables and derive the complete set of propagation and constraint equations
for the {\it kinematic}, {\it inhomogeneity} and {\it non\hs local}
quantities. In section IV we split these equations into {\it scalar},
{\it vector} and {\it tensor} parts, which we then analyse and discuss
in sections V\hs VII for the {\it low energy}, {\it high\hs energy}
and {\it dark radiation} dominated regimes respectively. Finally, in
section VIII we present our conclusions. For the most part we follow the
notation and convention of \cite{roy2,dunsbyetal}.
\section{Brane Dynamics}
\subsection{Geometric Formulation}
The implementation of the braneworld scenario considered
in~\cite{SMS} assumes that the whole spacetime is 5\hs D and governed
by the 5\hs D field equations $(A,B=0,...,4)$:
\begin{equation}
{G}^{(5)}_{AB}=-{\Lambda}_{(5)}{g}^{(5)}_{AB}
+{\kappa}_{(5)}^2\delta(\chi)[-\lambda g_{AB}+T_{AB}]\;. 
\label{eq:5d}
\end{equation}
These represent a 4\hs D brane at $\chi=0$ embedded in a vacuum bulk with
metric ${g}^{(5)}_{AB}$ and cosmological constant
${\Lambda}^{(5)}$; ${\kappa}^2_{(5)}$ is the 5\hs D
gravitational constant, $\lambda$ is the brane tension, ${g}_{AB}$
and $T_{AB}$ are respectively the metric and the energy-momentum
on the brane.  The 4\hs D field equations induced on the brane are
derived geometrically from (\ref{eq:5d}) assuming a $Z_2$ symmetry
with the brane at the fixed point, leading to modified Einstein
equations with new terms representing bulk effects:
\begin{equation}
G_{ab}=-\Lambda g_{ab}+\kappa^2
T_{ab}^{\rm tot}\,,
\label{2}
\end{equation}
where
\be
T_{ab}^{\rm tot}=T_{ab}+\sfrac{6}{\lambda}S_{ab} 
- \sfrac{1}{\kappa^2}E^{(5)}_{ab}\,.
\label{t-tot}
\ee
As usual  $\kappa^2=8\pi/M_{\rm p}^2$, and $(a,b=0,...,3)$.
The various physical constants and parameters appearing in the equations 
above are not independent, but related to each other by
\begin{equation}
\lambda=6\sfrac{\kappa^2}{\,\,\,\kappa_{(5)}^4} \,, ~~ \Lambda =
\sfrac{1}{2}[{\Lambda}_{(5)}+\kappa^2\lambda]\,.
\label{3}
\end{equation}
The tensor $S_{ab}$ represents non\hs linear matter corrections
given by
\be
S_{ab}=\sfrac{1}{12}T^c{}_cT_{ab}-\sfrac{1}{4}T_{ac}T^c{}_b
+\sfrac{1}{24}g_{ab}\left[3T_{cd} T^{cd}-\left(T^c{}_c\right)^2\right]\,.
\label{3'}
\ee
${E}^{(5)}_{ab}$ is the projection of the 5\hs D Weyl tensor $C^{(5)}_{ABCD}$
on to the brane: $E^{(5)}_{ab}=C^{(5)}_{ABCD}n^C n^Dg_a^Ag_b^B$, where $n_A$ is the
normal to the hypersurface $\chi =0$ ($n_A n^A =-1$).

Although the whole dynamics are 5\hs D and given by
(\ref{eq:5d}), from the 4\hs D point of view $E^{(5)}_{ab}$
is a non\hs local source term that carries bulk effects onto the
brane.

The energy\hs momentum tensor $T_{ab}$ is assumed to be conserved on the brane:
\be
\nabla^b T_{ab}=0\,,
\label{4}
\ee
and on using the 4\hs D contracted Bianchi identities
$\nabla^b G_{ab}=0$ an additional constraint is obtained:
\begin{equation}
\nabla^a{E}^{(5)}_{ab}=\sfrac{6\kappa^2}{\lambda}\nabla^b
S_{ab}\,, \label{5}
\end{equation}
which  shows how the non\hs local bulk effects are sourced by the
evolution and spatial inhomogeneity of the brane matter content.
\subsection{Cosmological Dynamics on the Brane}
In the following we describe the matter on the brane by a perfect
fluid with barotropic equation of state $p=(\gamma -1)\rho$. As usual,
we require $\gamma \geq 0$ to satisfy the dominant energy condition and
$\gamma\leq 2$ to preserve causality and therefore we restrict our
analysis to values of $0<\gamma\leq 2$. The case $\gamma =0 $ can be treated
similarly to our analysis below, but using different variables. We do not
study this special case here, but refer the reader to \cite{pmscalar} for
details on how to treat this case.

If $u^a$ is the matter 4\hs velocity and $h_{ab}=g_{ab}+u_a u_b$ projects
into the comoving rest space of a fundamental observer, the brane energy
momentum tensor is given by
\be
T_{ab}=\rho\,u_a u_b +ph_{ab}\;.
\ee
One can also decompose $E^{(5)}_{ab}$ in such a way that it is equivalent to a
trace-less energy momentum tensor with energy density $\rho^*$,
energy flux $q^*_{a}$ and anisotropic pressure $\pi^*_{ab}$ (see \cite{roy2}
for details):
\begin{equation}
-\sfrac{1}{\kappa^2}E^{(5)}_{ab}=\rho^*(u_a u_b+\sfrac{1}{3}h_{ab})
+q^*_{a}u_{b}+q^*_{b}u_{a}+\pi^*_{ab}\;.
\label{6}
\end{equation}
Since there is no evolution equation for the non\hs local anisotropic
pressure $\pi^*_{ab}$ we restrict our analysis to the case $\pi^*_{ab}=0$. 
This is dynamically justified in early time regimes \cite{roy2,CHL,Ruth}. 
Note in particular, that using (\ref{t-tot}) the total energy density
is given by
\be
\rho^{\rm tot}=\rho+\sfrac{1}{2\lambda}\rho^2+\rho^*\;.
\label{rho-tot}
\ee
In this way we can see that there are three essentially different 
energy regimes: when $\rho\gg\frac{\rho^2}{\lambda},\rho^*$ we recover 
general relativity (GR). When $\frac{\rho^2}{\lambda}\gg\rho,\rho^*$ 
we obtain the high energy limit, and when 
$\rho^*\gg\rho,\frac{\rho^2}{\lambda}$ we obtain the dark radiation
dominated regime.

Equation (\ref{4}) gives the usual energy and momentum conservation equations:
\begin{eqnarray}
&&\dot{\rho}+3\gamma H\rho=0\,,\label{e}\\
&&(\gamma-1)\D_a \rho +\gamma\rho A_a =0\,,\label{pc2}
\end{eqnarray}
where a dot denotes $u^b\nabla_b$, $H=3\D^a u_a$ is
the Hubble parameter of the background, $A_b=\dot{u}_b$ is the 
4\hs acceleration, and $\D_a$ denotes the spatially projected 
covariant derivative.

Using equations (\ref{4}) and (\ref{5}) we can obtain conservation
equations for the non\hs local quantities $\rho^*$ and $q^*_a$.
Restricting to linear perturbations of Robertson\hs Walker models
\cite{EB,pmscalar,BDE,DBE}, we obtain \footnote{Strictly
speaking, the variables defined in~\cite{EB} and those
defined in the same way in the brane context~\cite{roy2,GM,leong}
are 4-D, however they can easily be generalised to 5\hs D. Indeed
Bardeen like variables~\cite{bardeen,ks} have been defined in 5\hs D in
order to carry out a brane\hs bulk analysis (e.g.\ see~\cite{BMW}), however 
their relation to the covariant quantities used here has not yet been 
established (see \cite{BDE,DBE} for this relation in 4\hs D general 
relativity).}:
\begin{eqnarray}
&& \dot{\rho}^*+4H\rho^*+\D^a q^*_a  =0 \,, \label{lc1'} \\
&& \dot{q}^*_{a}+4Hq^*_a
+\sfrac{1}{3}\D_a \rho^*+\sfrac{4}{3}\rho^* A_a
=-\sfrac{\rho}{\lambda}\gamma\D_a \rho
\,.\label{lc2'}
\end{eqnarray}

Finally we note that the generalised Friedmann equation on the brane for a
flat, homogeneous isotropic background with vanishing 4-D cosmological
constant $\Lambda$ is
\begin{equation}\label{f}
H^2=\frac{\kappa^2\rho}{3} +\frac{\kappa^2\rho^2}{6\lambda}+
\frac{\kappa^2\rho^*}{3}\,.
\end{equation}
\section{Cosmological Perturbations}
In the following sections we will present a complete description
of general large\hs scale inhomogeneous perturbations for several different
flat ($K=0$) homogeneous isotropic background models with vanishing cosmological
constant on the brane ($\Lambda =0$).

Following the dynamical systems approach developed in 
\cite{dynamics1,dynamics2,dynamics3} and extended to the 
braneworld scenario in \cite{SC2,SC1} we define the dimensionless 
density parameter
\be
\Omega_\rho\equiv \frac{\kappa^2\rho}{3H^2}
\ee
as in general relativity, and
\be
\Omega_\lambda\equiv \frac{\kappa^2\rho^2}{6\lambda H^2},~~
\Omega_{\rho^*} \equiv \frac{\kappa^2 \rho^*}{3H^2}\nonumber
\ee
corresponding to the non\hs GR contributions to the Friedmann equation.
In this way we can classify the various background solutions by their
coordinates $(\Omega_\rho\,\Omega_\lambda,\Omega_{\rho^*})$ in the
phase space of Friedmann\hs Robertson\hs Walker (FRW) models.

The point $(1,0,0)$ corresponds to the flat GR Friedmann model
with $\lambda^{-1}=\rho^*=0$ and $a(t)\sim t^{2/{3\gamma}}$.
The point $(0,1,0)$ corresponds to the high energy model
${\cal F}_b$ with $\rho^*=0$ and $\rho\ll\rho^2/\lambda$; the scale
factor is given by $a(t)\sim t^{1/{3\gamma}}$, which can be found
by a limiting process \cite{coley2,langlois}. Finally the 
point $(0,0,1)$ corresponds to a model $(R)$ with scale factor 
evolution $a(t)\sim t^{1/2}$, satisfying $\rho=\lambda^{-1}=0$.

In what follows, we develop the perturbation equations for a
general background model $(\Omega_\rho\,\Omega_\lambda,\Omega_{\rho^*})$
and decompose the propagation and constraint equations into their respective scalar,
vector and tensor contributions in the usual way. We then evaluate these equations for
the three backgrounds described above and interpret the results.
\subsection{Dimensionless Variables}
The projected 4\hs D field equations (\ref{2}) can be covariantly
split using the Ricci identities and the Bianchi identities
\cite{roy2}. In the previous section we have already given the
conservation equations for energy and momentum (\ref{e}), (\ref{pc2})
and for the non\hs local energy density $\rho^*$ and flux $q^*_a$
(\ref{lc1'}), (\ref{lc2'}). The remaining equations correspond to
propagation and constraint equations for the kinematic quantities
i.e., the acceleration $A_b$, the vorticity $\omega_b$ and the shear
$\sigma_{ab}$, together with the electric and magnetic parts of the
Weyl tensor $E_{ab},~H_{ab}$ corresponding to the non-local gravitational
field on the brane.

Instead of using the standard quantities we define dimensionless
expansion normalised variables by
\be
W_{a} \equiv \frac{\omega_{a}}{H}\,,   ~
\Sigma_{ab} \equiv \frac{\sigma_{ab}}{H}\,,~
{\cal E}_{ab} \equiv \frac{E_{ab}}{H^2} \,,   ~
{\cal H}_{ab} \equiv \frac{H_{ab}}{H^2}\,.
\label{defdimless1}
\ee
We emphasise that $E_{ab}$ and the expansion normalised quantity 
${\cal E}_{ab}$ must not be confused with the 5\hs D Weyl tensor 
$E^{(5)}_{ab}$. It turns out that using the dimensionless vorticity variable
\be
W^*_a\equiv aH W_a
\label{defdimless2}
\ee
simplifies the calculations below; however, it should be noted
that $W_a$ and not $W^*_a$ is the physically relevant quantity.
We also use the dimensionless logarithmic time derivative $\tau$, 
defined by
\be
\{\}'=\frac{d}{d{\tau}}=\frac{d}{d~ln(a)}=\frac{1}{H}\frac{d}{dt}\;.
\nonumber
\ee

Density perturbations are physically characterised by the comoving
fractional density gradient defined by \cite{EB}:
\be
{\Delta}_a \equiv \frac{a}{\rho} \D_a \rho \,.~~
\label{definhom1}
\ee
In addition, it is convenient to define the following dimensionless gradients
describing inhomogeneity in the expansion rate $H$
and in the non\hs local energy density $\rho^*$ and flux $q^*_a$
\be
Z^*_a \equiv \frac{3a}{H}\D_a H\,,~
U^*_a \equiv \frac{\kappa^2 a}{H^2}\D_a{\rho^*}\,,
Q^*_a \equiv \frac{\kappa^2 a}{H}q^*_a\label{definhom2}\,.
\ee
Note that although $\Delta_a$ is not defined for the exact dark radiation
background $(R)$ where $\rho=0$, it is is well\hs defined in a neighbourhood
of $(R)$ and since $(R)$ is a saddle point in the phase space of the background
homogeneous models \cite{SC2}, this solution can never be exactly attained.
It is therefore sufficient to study arbitrarily small but non\hs zero
inhomogeneous perturbations of this model. In other words we may evaluate
the perturbation equations \emph{arbitrarily close} to the background
$(R)$, but not \emph{on} the exact background itself.

The above discussion suggests that it makes more sense to define specially tailored
inhomogeneity variables by normalising them with respect to the dominant energy density
term in the Friedmann equation.

For the low\hs energy limit we use the usual dimensionless perturbation
quantities :
\be
\Delta_a^{\rm (LE)}\equiv\frac{a}{\rho}\D_a{\rho},
~U_a^{\rm (LE)}\equiv \frac{a}{\rho} \D_a{\rho^*},  ~~
~Q_a^{\rm (LE)} \equiv \frac{1}{\rho} q^*_a\label{LE}\,,
\ee
however since $\rho_*$ is the dominant term in the {\it dark radiation}
dominated regime, the appropriate inhomogeneity variables are
\be
\Delta_a^{\rm (DE)}\equiv\frac{a}{\rho^*}\D_a{\rho},
~U_a^{\rm (DE)} \equiv \frac{a}{\rho^*} \D_a{\rho^*},  ~~
~Q_a^{\rm (DE)} \equiv \frac{1}{\rho^*} q^*_a\label{DE}\,.
\ee
Finally in the high\hs energy limit we define
\be
\Delta_a^{\rm (HE)}\equiv \frac{\lambda a}{\rho^2} \D_a{\rho}\,,
~U_a^{\rm (HE)}\equiv \frac{\lambda a}{\rho^2} \D_a{\rho^*}\,,  ~~
~Q_a^{\rm (HE)}\equiv \frac{\lambda}{\rho^2} q^*_a\label{HE}\,.
\ee
since the leading energy density term is proportional to $\rho^2/\lambda$,
which becomes our normalisation factor in this case.

When decomposing the equations into harmonics we will have to deal
with the curls of some of the variables. One approach would be to
eliminate the curls by deriving second and higher order
equations, or alternatively introduce new spatial harmonics
corresponding to the curls of the original harmonics. Instead, we
find it more convenient to define new variables corresponding to
the curls of the original quantities and derive propagation and
constraint equations for them. In this way we obtain a complete
closed set of linear differential equations which can be easily
solved. Note that {\em all} of the additional propagation and
constraint equations have to be satisfied, since these equations
are necessary to close the system.

We denote the curl of a quantity with an overbar and the key
variables of this type are:
\bea
\bar{W}^*_a \equiv \frac{1}{H}\curl{W^*_a}\,,
\bar{\Sigma}_{ab} \equiv
\frac{1}{H}\curl{\Sigma_{ab}}\,,   \nonumber\\
\bar{{\cal E}}_{ab} \equiv \frac{1}{H}\curl{\cal E}_{ab} \,,
\bar{\cal H}_{ab} \equiv \frac{1}{H}\curl{\cal H}_{ab}
\label{defbar1}
\eea
 and
\be
\bar{Q}^*_a \equiv
\frac{1}{H}\curl{Q^*_a}\label{defbar2}\,.
\ee
\subsection{Dimensionless Linearized Propagation and Constraint Equations}
The complete set of propagation and constraint equations for the
kinematic and non\hs local quantities on the brane were developed
in \cite{roy2}. Here we extend this work by presenting the complete set
of evolution and constraint equations for the dimensionless variables defined
by (\ref{defdimless1}-\ref{defdimless2}) and (\ref{defbar1}).

We begin with the generalised Raychaudhuri equation
\be
-\frac{H'}{H}=(1+q)+\sfrac{\gamma-1}{\gamma}\sfrac{1}{3aH^2}\D^a\Delta_a ,\label{friedman}
\ee
where 
\bea
q=\sfrac{3}{2}\gamma\Omega_\rho+3\gamma\Omega_\lambda+2\Omega_{\rho^*}-1
\nonumber
\eea
is the usual deceleration parameter.

The remaining propagation equations are given by
\bea
&&{W^*_a}'=(3\gamma-4)W^*_a\nonumber\,,\\
&&\Sigma_{ab}'=(q-1)\Sigma_{ab} -{\cal E}_{ab}-\sfrac{\gamma-1}{\gamma}(aH)^{-2}
\D_{<a} {\Delta}_{b>}\nonumber\,,\\
&&{\cal E}_{ab}'=(2q-1){\cal E}_{ab}-(q+1)\Sigma_{ab}\nonumber\\
&&~~~~-\sfrac{1}{2aH^2}\D_{<a} {Q^*}_{b>}+\bar{\cal H}_{ab}\nonumber\,,\\
&&{\cal H}_{ab}'=(2q-1){\cal H}_{ab}-\bar{\cal E}_{ab}\,,
\label{prop}
\eea
which are subject to following dimensionless constraints:
\bea
&&a\D^b W^*_b=0\nonumber\,,\\
&&a\D^b\Sigma_{ab}=\bar{W}^*_a+\sfrac{2}{3}Z^*_a-Q^*_a\,,\nonumber\\
&&H\bar{\Sigma}_{ab}=-\sfrac{1}{aH}\D_{<a}W^*_{b >}+H{\cal H}_{ab}\nonumber\,,\\
&&a\D ^b {\cal E}_{ab}=(\Omega_\rho +2\Omega_\lambda)\Delta_a
+\sfrac{1}{3} U^*_{a}-Q^*_a\nonumber\,,\\
&&a\D ^b{\cal H}_{ab}=2(q+1)W^*_a -\sfrac{1}{2}\bar{Q}^*_a\,,
\label{constr}
\eea
where the angle bracket is defined by
\be
D_{<a}W^*_{b>}\equiv D_{(a}W^*_{b)}-\sfrac{1}{3}D^cW^*_ch_{ab}\;.
\nonumber
\ee
Equations for the inhomogeneity variables $\Delta_a$, $Z^*_a$, $Q^*_a$ and $Q^*_a$.
are given by
\bea
&&{\Delta_a}'=(3\gamma-3)\Delta_a-\gamma{Z}^*_a\nonumber\,,\\
&&{{Z}^*_a}'=(q-1){Z}^*_a-\sfrac{3}{2}[\,\Omega_\rho\!+\!(6\gamma\!+\!2)\Omega_\lambda
\!-\!\sfrac{4\gamma\!-\!4}{\gamma}\Omega_{\rho^*}\,]\,\Delta_a\nonumber\\
&&~~~~-{U}^*_a-6(\gamma-1)\bar{W}_a-\sfrac{\gamma-1}{\gamma}\sfrac{1}{H^2}
\D^2\Delta_a\,,\nonumber\\
&&{Q^*_a}'=(q-2)Q^*_a -\sfrac{1}{3}U^*_a
+(4\sfrac{\gamma-1}{\gamma}\Omega_{\rho^*}-6\gamma \Omega_\lambda) \Delta_a\nonumber\,,\\
&&{{U}^*_a}'=(2q-2){U}^*_a-4\Omega_{\rho^*}\,Z^*_a\nonumber\\
&&~~~~+\sfrac{12\gamma-12}{\gamma}\Omega_{\rho^*}\,\Delta_a-\sfrac{1}{H^2}\D_a(\D^b {Q}^*_b)\,.
\label{inhomprop}
\eea
Finally, using the definitions (\ref{defbar1}) and (\ref{defbar2}),
we obtain equations for the curls of the original quantities:
\bea
&&{\bar{W}}^*{}_a'=(q+3\gamma -4)\bar{W}^*_a\nonumber\,,\\
&&{\bar{\Sigma}_{ab}}'=(2q-1)\bar{\Sigma}_{ab}-\bar{\cal E}_{ab}\nonumber\,,\\
&&{\bar{\cal E}_{ab}}'=(3q-1)\bar{\cal E}_{ab}
-(q+1)\bar\Sigma_{ab}+\sfrac{3}{2H^2}\D_{<a}\D^c
{\cal H}_{b>c}\nonumber\\
&&~~~~-\sfrac{1}{4aH^2}\D_{<a}\bar{Q}^*_{b >}
-\sfrac{1}{H^2}\D^2 {\cal H}_{ab} \nonumber\,,\\
&&{\bar{\cal H}_{ab}}'=(3q-1)\bar{\cal H}_{ab}-\sfrac{3}{2H^2}
\D_{<a}\D^c{\cal E}_{b>c}
+\sfrac{1}{H^2}\D^2{\cal E}_{ab} \nonumber\,,\\
&&{\bar{Q}}^*{}_a'=(2q-2)\bar{Q}^*_a+4[(6\gamma-8)\Omega_{\rho^*}
-9\gamma^2\Omega_\lambda]{W}^*_a\,,  
\label{curlprop}
\eea
which are subject to the following constraints (obtained by taking the curls of (\ref{constr})):
\bea
&&a\D^b \bar{W}_b^*=0 \nonumber\,,\\
&&a\D^b \bar{\Sigma}_{ab}=2(q+1){W}^*_a
-\sfrac{1}{2}\bar{Q}^*_a-\sfrac{1}{2H^2}\D^2{W}^*_a\nonumber\,,\\
&&\bar{\cal H}_{ab}=\sfrac{1}{2aH}\D_{<a}\bar{W}^*_{b>}
-\sfrac{1}{H^2}\D^2{\Sigma}_{ab}+\sfrac{3}{2H^2}\D_{<a}\D^c\Sigma_{b>c}\nonumber\,,\\
&&a\D^b \bar{\cal E}_{ab}=2(q+1){W}^*_a
-\sfrac{1}{2}\bar{Q}^*_a\nonumber\,,\\
&&a\D^b \bar{\cal H}_{ab}=(q+1)\bar{W}^*_a+
\sfrac{1}{4H^2}(\D^2{Q}^*_a -\D_a (\D^b{Q}^*_b))\nonumber\,,\\
&&a\D^a \bar{Q}^*_a =0\label{curlconstr}\,.
\eea
\section{Harmonic Decomposition}
In order to solve these equation we employ the standard approach 
of expanding the variables in these equations in terms of
scalar (S), vector (V) and tensor (T) harmonics $Q$ \footnote{The properties
of these harmonics are discussed extensively in the appendix of \cite{BDE}.}. 
These harmonics are eigenfunctions of the covariantly defined 
Laplace\hs Beltrami operator \cite{BDE}:
\be
\D^2 Q\equiv\D_a\D^a Q=-\sfrac{k^2}{a^2} Q\;, 
\ee
where $k$ is the wave number corresponding to a comoving scale 
$\lambda\equiv 2\pi a/k$. This yields a covariant and gauge invariant 
splitting into three sets of evolution and constraint equations for 
scalar, vector and tensor modes.

Thus a scalar $X$, vector $X^a$ (orthogonal to $u^a$) and tensor $X^{ab}$
(orthogonal to $u^a$) can be expanded as follows
\bea
&&X=X^S Q^S \nonumber\,,\\
&&X_a=k^{-1}X^S Q_{a}^S+X^V Q_{a}^V \nonumber\,,\\
&&X_{ab}=k^{-2}X^S Q_{ab}^S+k^{-1}X^VQ_{ab}^V+X^T Q_{ab}^T\;.
\eea
In what follows we drop the subscripts $S,~V,~T$ and also restrict
our analysis to the long wavelength limit defined by $\sfrac{k^2}{a^2H^2}<<1$.
\subsection{Scalar Perturbations}
In the long wavelength limit the scalar evolution equations for the
kinematics (which follow after expanding (\ref{prop}) in terms of scalar
harmonics) are given by
\bea
&&\Sigma '=(q-1)\Sigma -{\cal E}\;,\nonumber \\
&&{\cal E}'=(2q-1){\cal E}-(q+1)\Sigma\;,
\eea
and these are subject to the following constraints 
(which follow from (\ref{constr}):
\bea
&&W=\bar{W}={\cal H}=\bar{\cal H}=\bar{\Sigma}=\bar{\cal E}=\bar{Q}=0\;, \nonumber \\
&&2{Z_*}=2\Sigma +3{Q_*}\;, \nonumber \\
&&3(\Omega_\rho +2\Omega_\lambda)\Delta = 2{\cal E}+3{Q_*}-{U_*}\;.
\label{cs2}
\eea
In addition, we have scalar evolution equations for the
inhomogeneity variables (which follow from (\ref{inhomprop})
after harmonic analysis):
\bea
&&{\Delta}'=(3\gamma-3)\Delta-\gamma{Z_*}\;, \nonumber \\
&&Z_* '=(q-1){Z}_*-[\sfrac{3}{2}\Omega_\rho+(9\gamma+3)\Omega_\lambda\\
&&~~~~-6\sfrac{\gamma-1}{\gamma}\Omega_{\rho^*}\,]\,\Delta -{U}_*\;, \nonumber \\
&&Q_* ' =(q-2){Q_*}
-\sfrac{1}{3}U_*+(4\sfrac{\gamma-1}{\gamma}\Omega_{\rho^*}-
6\gamma\Omega_\lambda)\,\Delta\label{Q's}\;, \nonumber \\
&&U_* '=(2q-2){U_*}-4\Omega_{\rho^*}\,Z_*
+12\sfrac{\gamma-1}{\gamma}\Omega_{\rho^*}\,\Delta\;.
\label{U's}
\eea
\subsection{Vector Perturbations}
Expanding equations (\ref{prop}) and (\ref{curlprop}) in terms
of vector harmonics, we obtain the following evolution
equations for the kinematic and non\hs local quantities
together with their curls:
\bea
&&{W_*}'=(3\gamma-4){W_*}\;,\nonumber \\
&&\Sigma '=(q-1)\Sigma -{\cal E} -2(\gamma-1)\bar{W}_*\;,\nonumber \\
&&{\cal E}'=(2q-1){\cal E}-(q+1)\Sigma+\bar{\cal H}\;,\nonumber \\
&&{\cal H}'=(2q-1){\cal H}-\bar{\cal E}\;,\nonumber \\
&&{\bar{W}_*}'=(3\gamma-4+q)\bar{W_*}\;,\nonumber \\
&&{\bar{\Sigma}}'=(2q-1)\bar{\Sigma}-\bar{\cal E}\;,\nonumber \\
&&{\bar{\cal E}}'=(3q-1)\bar{\cal E}-(q+1)\bar{\Sigma}\;,\nonumber \\
&&{\bar{\cal H}}'=(3q-1)\bar{\cal H}\;,\nonumber \\
&&{\bar{Q}_*}'=(2q-2)\bar{Q}_*+4[(6\gamma-8)\Omega_{\rho^*}
-9\gamma^2\Omega_\lambda]{W}_*\,.\nonumber
\label{vectorprop}
\eea
These equations are subject to the following constraints,
which are obtained from (\ref{constr}) and  (\ref{curlconstr}):
\bea
&&4{Z_*}=3\Sigma+6Q_*-6\bar{W}_*\;, \nonumber \\
&&6(\Omega_\rho +2\Omega_\lambda)\Delta =3{\cal E}+6Q_*-2U_*\;, \nonumber \\
&&\bar{\Sigma}={\cal H}\;, \nonumber \\
&&{\cal H}=4(q+1){W_*}-\bar{Q_*}\;\nonumber \\.
&&\bar{\cal H}=2(q+1) \bar{W}_*=0\;,\nonumber \\
&&\bar{\Sigma}=4(q+1){W_*}-\bar{Q}_*\;,\nonumber \\
&&\bar{\cal E}=4(q+1){W_*}-\bar{Q}_*\;.
\eea
Equations for the vector parts of the inhomogeneity variables
follow from (\ref{inhomprop}) and are given by
\bea
&&{\Delta}'=(3\gamma-3)\Delta-\gamma{Z_*}\;, \nonumber \\
&&{Z_*}'=(q-1){Z}_*-[\sfrac{3}{2}\Omega_\rho+3(3\gamma+1)\Omega_\lambda
-6\sfrac{\gamma-1}{\gamma}\Omega_{\rho^*}\,]\,\Delta\nonumber\\
&&-{U}_*-6(\gamma-1)\bar{W}_*\;, \nonumber \\
&&{Q_*}'=(q-2){Q_*}-\sfrac{1}{3}{U_*}+(4\sfrac{\gamma-1}
{\gamma}\Omega_{\rho^*}-6\gamma\Omega_\lambda)\Delta\;, \nonumber \\
&&{U_*}'=(2q-2){U}_*-4\Omega_{\rho^*}\,Z_*+12
\sfrac{\gamma-1}{\gamma}\Omega_{\rho^*}\,\Delta\;. \label{U'v}
\eea
\subsection{Tensor Perturbations}
Finally the long wavelength behaviour of tensor perturbations are
obtained by expanding (\ref{prop}) and (\ref{curlprop}) in terms of 
tensor harmonics:
\bea
&&\Sigma '=(q-1)\Sigma -{\cal E}\;,\nonumber \\
&&{\cal E}'=(2q-1){\cal E}-(q+1)\Sigma\;,\nonumber \\
&&{\cal H}'=(2q-1){\cal H}-\bar{\cal E}\;,\nonumber \\
&&{\bar{\Sigma}}'=(2q-1)\bar{\Sigma} -\bar{\cal E}\;,\nonumber \\
&&{\bar{\cal E}}'=(3q-1)\bar{\cal E}-(q+1)\bar{\Sigma}\;, \label{pe}
\eea
subject to the following constraints
\be
\bar{\cal H}=0\;,~~
\bar{\Sigma}={\cal H}\;.
\ee
\section{Low Energy Limit: The GR Background}
We begin with perturbations in the low\hs energy limit,
defined by
$\rho\gg \rho^2/\lambda$ and $\rho\gg\rho^*$ 
or $\Omega_\rho \gg \Omega_\lambda$ 
and $\Omega_\rho \gg \Omega_{\rho^*}$. 
We therefore evaluate the perturbation equations 
(\ref{cs2}-\ref{pe}) in the limit
$(\Omega_\rho,\Omega_\lambda,\Omega_{\rho^*})\rightarrow(1,0,0)$.
Using the energy conservation equation
(\ref{e}), the Friedmann equation (\ref{f}) can be solved to give the
background scale factor $a$ and the Hubble parameter $H$:
\be
a(t)=\left(t/t_0\right)^{2/3\gamma},
~~H=H_0a^{-3\gamma/2}\;,
\ee
where we fix an arbitrary initial condition by choosing  $a_0=a(t_0)=1$.
The deceleration parameter is given by
\be
q=\sfrac{3}{2}\gamma-1\,,
\ee
and, as usual,
\be
\rho=\rho_0a^{-3\gamma}\;,
\ee
where $H_0^2=\frac{\kappa^2}{3}\rho_0$.
\subsection{Scalar Perturbations}
In this case the propagation equations for the
kinematic quantities are
\bea
&&\Sigma '=(\sfrac{3}{2}\gamma-2)\Sigma -{\cal E}\,,\nonumber\\
&&{\cal E}'=(3\gamma-3){\cal E}-\sfrac{3}{2}\gamma\Sigma\,,
\eea
while the constraints are given by
\bea
&&W=\bar{W}={\cal H}=\bar{\cal H}=\bar{\Sigma}=\bar{\cal E}=\bar{Q}=0\;,\nonumber \\
&&2{Z_*}=2\Sigma +3{Q_*}\,,\nonumber \\
&&3\Delta=2{\cal E}+3{Q_*}-{U_*}\;,
\eea
and the equations for the inhomogeneity variables become
\bea
&&{\Delta}'=(3\gamma-3)\Delta-\gamma{Z_*}\,,\nonumber \\
&&{Z_*}'=(\sfrac{3}{2}\gamma-2){Z}_*-\sfrac{3}{2}\Delta -{U}_*\,,\nonumber \\
&&{Q_*}'=(\sfrac{3}{2}\gamma-3){Q_*}-\sfrac{1}{3}U_*\,,\nonumber \\
&&{U_*}'=(3\gamma-4){U_*}\;.
\eea
The above equations can be easily solved to give
\bea
&&\Sigma=\Sigma_0 a^{3\gamma-2}+\Sigma_1 a^{\sfrac{3}{2}\gamma-3}\;,\nonumber \\
&&{\cal E}=-\sfrac{3}{2}\gamma\Sigma_0a^{3\gamma-2}+\Sigma_1a^{\sfrac{3}{2}\gamma-3}\;,
\eea
and
\bea
&&{\Delta}=-\gamma\Sigma_0a^{3\gamma-2}+(\sfrac{2}{3}\Sigma_1+Q^*_0)
a^{\sfrac{3}{2}\gamma-3}-\sfrac{\gamma}{3\gamma-2}U^*_0a^{3\gamma-4}\,,\nonumber \\
&&{Z}_*=\Sigma_0a^{3\gamma-2}+(\Sigma_1+\sfrac{3}{2}Q^*_0)a^{\sfrac{3}{2}
\gamma-3}-\sfrac{1}{3\gamma-2}U^*_0 a^{3\gamma-4}\,,\nonumber \\
&&{Q}_*=Q^*_0a^{\sfrac{3}{2}\gamma-3}-\sfrac{2}{3}\sfrac{1}{3
\gamma-2}U^*_0a^{3\gamma-4}\,,\nonumber \\
&&{U}_*=U^*_0 a^{3\gamma-4}
\eea
for $\gamma\neq\sfrac{2}{3}$, and
\bea
&&{\Delta}=-\sfrac{2}{3}\Sigma_0+( \sfrac{2}{3}\Sigma_1+Q^*_0
-\sfrac{1}{3}U^*_0)a^{-2}-\sfrac{1}{3}U^*_0\ln a\,\, a^{-2}\,,\nonumber \\
&&{Z}_*=\Sigma_0+(\Sigma_1+\sfrac{3}{2}Q^*_0)a^{-2}
-\sfrac{1}{2}U^*_0\ln a\,\, a^{-2}\,,\nonumber \\
&&{Q}_*=Q^*_0 a^{-2}-\sfrac{1}{3}U^*_0\ln a\,\, a^{-2}\,,\nonumber \\
&&{U}_*=U^*_0 a^{-2}
\eea
if $\gamma=\sfrac{2}{3}$.

Here $\Sigma_0,~\Sigma_1,~Q^*_0,~U^*_0$ are arbitrary
constants of integration, corresponding to the four
independent modes.

Finally the solutions for $Q_*,~U_*$ can be converted
into the the physical quantities $Q^{\rm (LE)},~U^{\rm (LE)}$
(which correspond to the scalar modes of $Q_a,~U_a$ defined
in (\ref{LE})) using the background solutions for $H$ and $\rho$:
\bea
&&Q^{\rm (LE)}=\sfrac{1}{3H_0}Q^*_0a^{3\gamma-4}
-\sfrac{2}{9H_0}\sfrac{1}{3\gamma-2}U^*_0a^{\frac{9}{2}\gamma-5}\;,\nonumber \\
&&U^{\rm (LE)}=\sfrac{1}{3}U^*_0 a^{3\gamma-4}
\eea
for $\gamma\neq\sfrac{2}{3}$ and
\bea
&&Q^{\rm (LE)}=\sfrac{1}{3H_0}Q^*_0 a^{-2}-\sfrac{1}{9H_0}U^*_0\ln a\,
a^{-2}\;,\nonumber \\
&&U^{\rm (LE)}=\sfrac{1}{3}U^*_0 a^{-2}\;.
\eea
if $\gamma=\sfrac{2}{3}$.
\begin{table*}
\caption{\label{tab:table1} Large scale contributions of the
different modes to the geometric and kinematic quantities for the
low energy background. We assume $0<\gamma\leq 2$,
$\gamma\neq\sfrac{2}{3}$, and we omit non\hs zero constant coefficients. The
first line in this table should be read as
$\Sigma=\alpha\Sigma_0 a^{3\gamma-2}+\beta\Sigma_1 a^{\frac{3}{2}\gamma-3}$,
where $\alpha,\beta$ are some non\hs zero constants, for all scalar, vector
and tensor modes. General Relativity is recovered when $Q^*_0=U^*_0=0$.}
\begin{tabular}{c|cccc|cccc|cccc}
&  &  &  &  &  &  &  &  &  &  &  &\\
harmonic& &scalar&& &&vector& & & &tensor & \\
\hline
&  &  &  &  &  &  &  &  &  &  &  &\\
mode& $a^{3\gamma-2}$ & $a^{\sfrac{3}{2}\gamma-3}$ & $a^{3\gamma-4}$ &$a^{\sfrac{9}{2}\gamma-5}$&
$a^{3\gamma-2}$ &
$a^{\sfrac{3}{2}\gamma-3}$ & $a^{3\gamma-4}$& $a^{\sfrac{9}{2}\gamma-5}$ &$a^{3\gamma-2}$ &
$a^{\sfrac{3}{2}\gamma-3}$ & $a^{3\gamma-4}$ & $a^{\sfrac{9}{2}\gamma-3}$ \\
\hline
&  &  &  &  &  &  &  &  &  &  &  &  \\
$\Sigma$ & $\Sigma_0$ & $\Sigma_1$ & -&- & $\Sigma_0$ & $\Sigma_1$ & - & - & $\Sigma_0$
& $\Sigma_1$ & - & - \\
${\cal E}$ & $\Sigma_0$ & $\Sigma_1$ & -& - & $\Sigma_0$ & $\Sigma_1$ & - & -
& $\Sigma_0$ & $\Sigma_1$ & - & - \\
${\cal H}$ & - & -&  - & - & - & - & ${\cal H}_0$ &  - & - & - & ${\cal H}_0$ & ${\cal H}_1$ \\
$W$ & - & - & - & - & -& - & - & $ W^*_0$ & - & - & - & - \\
$\Delta$ & $\Sigma_0$ & $2\Sigma_1+3Q^*_0$ & $U^*_0$ &-&
$\Sigma_0$
& $\Sigma_1+2Q^*_0$ & $U^*_0$ & - & - & - & - & - \\
$Z^*$ & $\Sigma_0$ & $2\Sigma_1+3Q^*_0$ & $U^*_0$ &-& $\Sigma_0$ &
$\Sigma_1+2Q^*_0$& $U^*_0$ & - & - & - & - & - \\
$Q^{\rm (LE)}$ &-& - & $Q^*_0$ & $U^*_0$ & -&- & $Q^*_0$ & $U^*_0$
& - & - & - & - \\
$U^{\rm (LE)}$ & - & - & $U^*_0$ & - &-& - & $U^*_0$ & - & - & - & - & - \\
\end{tabular}
\end{table*}
\subsection{Vector Perturbations}
For vector perturbations the complete set of propagation equations
for the kinematic and non\hs local quantities are given by the
ten\hs dimensional system
\bea
&&{W_*}'=(3\gamma-4){W_*}\;,\nonumber \\
&&\Sigma '=(\sfrac{3}{2}\gamma-2)\Sigma -{\cal E}\;,\nonumber \\
&&{\cal E}'=(3\gamma-3){\cal E}-\sfrac{3}{2}\gamma\Sigma\;,\nonumber \\
&&{\cal H}'=(3\gamma-3){\cal H}-\bar{\cal E}\;,\nonumber \\
&&{\bar{\Sigma}}'=(3\gamma-3)\bar{\Sigma}-\bar{\cal E}\;,\nonumber \\
&&{\bar{\cal E}}'=(\sfrac{9}{2}\gamma-4)\bar{\cal E}
-\sfrac{3}{2}\gamma\bar{\Sigma}\;,\nonumber \\
&&{\bar{Q}_*}'=(3\gamma-4)\bar{Q}_*\;.
\eea
subject to the following constraints
\bea
&&4{Z_*}=3\Sigma+6{Q_*}\;,\nonumber \\
&&6\Delta=3{\cal E}+6{Q_*}-2U_*\;,\nonumber \\
&&{\cal H}=\bar{\Sigma}=\bar{\cal E}=6\gamma{W_*}-\bar{Q}_*\;,\nonumber \\
&&\bar{\cal H}=\bar{W}_*=0\;,
\eea
while the propagation equations for the inhomogeneity variables are
\bea
&&{\Delta}'=(3\gamma-3)\Delta-\gamma{Z_*}\;,\nonumber \\
&&{Z_*}'=(\sfrac{3}{2}\gamma-2){Z}_*-\sfrac{3}{2}\Delta-{U}_*\;,\nonumber \\
&&{Q_*}'=(\sfrac{3}{2}\gamma-3){Q_*}-\sfrac{1}{3}{U_*}\;,\nonumber \\
&&{U_*}'=(3\gamma-4){U}_*\;.
\eea
Solutions can again be easily obtained and are given by
\bea
&&\Sigma=\Sigma_0 a^{3\gamma-2}+\Sigma_1 a^{\sfrac{3}{2}\gamma-3}\,,\nonumber \\
&&{\cal E}=-\sfrac{3}{2}\gamma\Sigma_0a^{3\gamma-2}
+\Sigma_1a^{\sfrac{3}{2}\gamma-3}\,,\nonumber \\
&&{\cal H}=\bar{\Sigma}=\bar{\cal E}={\cal H}_0 a^{3\gamma-4}\,,\nonumber \\
&&W=(aH)^{-1}W_*=\sfrac{3}{2}\gamma W^*_0 a^{\sfrac{9}{2}\gamma-5}\,,\nonumber \\
&&\bar{Q}_*=(6\gamma W^*_0-{\cal H}_0) a^{3\gamma-4}\,,
\eea
and
\bea
&&{\Delta}=-\sfrac{3}{4}\gamma\Sigma_0a^{3\gamma-2}
+(\sfrac{1}{2}\Sigma_1+Q^*_0)a^{\sfrac{3}{2}\gamma-3}
-\sfrac{\gamma}{3\gamma-2}U^*_0a^{3\gamma-4}\,,\nonumber \\
&&{Z}_*=\sfrac{3}{4}\Sigma_0a^{3\gamma-2}
+\sfrac{3}{4}(\Sigma_1+2 Q^*_0)a^{\sfrac{3}{2}\gamma-3}
-\sfrac{1}{3\gamma-2}U^*_0 a^{3\gamma-4}\,,\nonumber \\
&&Q^{\rm (LE)}=\sfrac{1}{3H_0}Q^*_0a^{3\gamma-4}
-\sfrac{2}{9H_0}\sfrac{1}{3\gamma-2}U^*_0a^{\frac{9}{2}\gamma-5}\,,\nonumber \\
&&U^{\rm (LE)}=\sfrac{1}{3}U^*_0 a^{3\gamma-4}
\eea
for $\gamma\neq\sfrac{2}{3}$, and
\bea
&&{\Delta}=-\sfrac{1}{2}\Sigma_0 +(\sfrac{1}{2}\Sigma_1+Q^*_0-\sfrac{1}{3}U^*_0)a^{-2}
-\sfrac{1}{3}U^*_0\ln a\,\, a^{-2}\,,\nonumber \\
&&{Z}_*=\sfrac{3}{4}\Sigma_0+(\sfrac{3}{4}\Sigma_1+\sfrac{3}{2}Q^*_0)a^{-2}
-\sfrac{1}{2}U^*_0\ln a\,\, a^{-2}\,,\nonumber \\
&&Q^{\rm (LE)}=\sfrac{1}{3H_0}Q^*_0 a^{-2}-\sfrac{1}{9H_0}U^*_0\ln a\,\, a^{-2}\,,\nonumber \\
&&U^{\rm (LE)}=\sfrac{1}{3}U^*_0 a^{-2}
\eea
if $\gamma=\sfrac{2}{3}$.
Again, we have converted the solutions for $Q_*,~U_*$ into
the physical quantities $Q^{\rm (LE)},~U^{\rm (LE)}$.

There are six independent modes corresponding to the constants of integration
$\Sigma_0,~\Sigma_1,~W_0,~{\cal H}_0,~Q^*_0$ and $U^*_0$.
\subsection{Tensor Perturbations}
For tensor perturbations the propagation equations in the long
wavelength limit are
\bea
&&\Sigma '=(\sfrac{3}{2}\gamma-2)\Sigma -{\cal E}\,,\nonumber \\
&&{\cal E}'=(3\gamma-3){\cal E}-\sfrac{3}{2}\gamma\Sigma\,,\nonumber \\
&&{\cal H}'=(3\gamma-3){\cal H}-\bar{\cal E}\,,\nonumber \\
&&{\bar{\Sigma}}'=(3\gamma-3)\bar{\Sigma}-\bar{\cal E}\,,\nonumber \\
&&{\bar{\cal E}}'=(\sfrac{9}{2}\gamma-4)\bar{\cal E}-\sfrac{3}{2}\gamma\bar{\Sigma}\,,
\eea
subject to the following constraints:
\be
\bar{\cal H}=0,~~\bar{\Sigma}={\cal H}\,.
\ee
The solutions are
\bea
&&\Sigma=\Sigma_0 a^{3\gamma-2}+\Sigma_1 a^{\sfrac{3}{2}\gamma-3}\,,\nonumber \\
&&{\cal E}=-\sfrac{3}{2}\gamma\Sigma_0a^{3\gamma-2}+\Sigma_1a^{\sfrac{3}{2}\gamma-3}\,,\nonumber \\
&&{\cal H}=\bar{\Sigma}={\cal H}_0 a^{3\gamma-4}+{\cal H}_1 a^{\sfrac{9}{2}\gamma-3}\,,\nonumber \\
&&\bar{\cal E}={\cal H}_0 a^{3\gamma-4}
-\sfrac{3}{2}\gamma{\cal H}_1a^{\sfrac{9}{2}\gamma-3}\,,
\eea
where $\Sigma_0,~\Sigma_1,~{\cal H}_0,~{\cal H}_1$ are four
independent constants of integration.
\section{The Dark Energy Era}
The dark energy dominated regime is characterised by
$\rho^*\gg\rho$ and $\rho^*\gg\rho^2/\lambda$ 
or $\Omega_{\rho^*}\gg \Omega_\rho$ and $\Omega_{\rho^*}\gg \Omega_\lambda$, 
so we now evaluate the perturbation equations in the limit 
$(\Omega_\rho,\Omega_\lambda,\Omega_{\rho^*})\rightarrow(0,0,1)$.

The background solution $(R)$, has the same metric as a flat
radiation FRW model, with $\rho=0$ and $\rho^*=\rho^*_0 a^{-4}$.
The scale factor $a$, Hubble parameter $H$ and
deceleration parameter $q$ are given by
\be
a(t)=\left(t/t_0\right)^{1/2},
~~H=H_0 a^{-2},~~q=1\;.
\ee
As explained in section III the perturbation equations are only defined
for small but non\hs zero energy density $\rho$, or equivalently
arbitrarily close but not {\em on} the exact background $(R)$. We
therefore use $\rho=\rho_0 a^{-3\gamma}$ for small but
non\hs zero values of $\rho_0$.
\subsection{Scalar Perturbations}
In the case of scalar perturbations the propagation equations
reduce to:
\bea
&&\Sigma '= -{\cal E}\,,\\
&&{\cal E}'={\cal E}-2\Sigma\;,
\eea
subject to the following constraints:
\bea
&&W=\bar{W}={\cal H}=\bar{\cal H}=\bar{\Sigma}=\bar{\cal E}=\bar{Q}=0\;,\nonumber\\
&&2\Sigma +3{Q_*}=2{Z_*}\,,\nonumber \\
&&2{\cal E}+3{Q_*}-{U_*}=3(\Omega_\rho+2\Omega_\lambda)\Delta\;,
\eea
while the equations for the inhomogeneity variables are
\bea
&&{\Delta}'=(3\gamma-3)\Delta-\gamma{Z_*}\;,\nonumber \\
&&{Z_*}'=6\sfrac{\gamma-1}{\gamma}\Delta -{U}_*\;,\nonumber \\
&&{Q_*}' =-{Q_*}-\sfrac{1}{3}U_*+4\sfrac{\gamma-1}{\gamma}\Delta\;,\nonumber \\
&&{U_*}'=-4Z_*+12\sfrac{\gamma-1}{\gamma}\Delta\;.
\eea
The following solutions can then be obtained:
\bea
&&\Sigma=\Sigma_0 a^2+\Sigma_1 a^{-1}\;,\nonumber \\
&&{\cal E}=-2\Sigma_0a^2+\Sigma_1a^{-1}\;,
\eea
and
\bea
&&\Delta=\Delta_0+ \Delta_1 a^{3\gamma -5}+
\sfrac{3\gamma}{3\gamma-7}\Sigma_0 a^2\,,\nonumber \\
&&Z_*=\sfrac{3\gamma-3}{\gamma}\Delta_0+ \sfrac{2}{\gamma}\Delta_1 a^{3\gamma -5}
+3\sfrac{3\gamma-5}{3\gamma-7}\Sigma_0 a^2\,,\nonumber \\
&&Q_*=\sfrac{2\gamma-2}{\gamma}\Delta_0+ \sfrac{4}{3\gamma}\Delta_1 a^{3\gamma -5}
+\sfrac{4}{3}\sfrac{3\gamma-4}{3\gamma-7}\Sigma_0 a^2-\sfrac{2}{3}\Sigma_1 a^{-1}\,,\nonumber \\
&&U_*=\sfrac{6\gamma-6}{\gamma}\Delta_0+ \sfrac{4}{\gamma}\Delta_1 a^{3\gamma -5}
+\sfrac{12}{3\gamma-7}\Sigma_0 a^2
\eea
for $\gamma\neq \sfrac{7}{3}$. We do not give the solutions for
$\gamma=\sfrac{7}{3}$, since all values of $\gamma>2$ are outside
the region of interest.

There are four constants of integration
$\Sigma_0,~\Sigma_1,~\Delta_0,~\Delta_1$
corresponding to the four independent modes.

The scalar contributions to the physical quantities
defined in (\ref{DE}) can then easily be obtained:
\bea
&&Q^{\rm (DE)}=\sfrac{2\gamma-2}{3H_0\gamma}\Delta_0 a
+ \sfrac{4}{9H_0\gamma}\Delta_1 a^{3\gamma -4}
+\sfrac{4}{9H_0}\sfrac{3\gamma-4}{3\gamma-7}\Sigma_0 a^3\nonumber \\
&&~~~~~~~~~~-\sfrac{2}{9H_0}\Sigma_1\,,\nonumber \\
&&U^{\rm (DE)}=U_*\;.
\eea
In particular, the density perturbation $\Delta^{\rm (DE)}$
can be written as $\Omega_{\rho^*}\Delta^{\rm (DE)}=\Omega_\rho \Delta$,
and using the fact that $\Omega_\rho,\Omega_\lambda\geq 0$, we find that
$\Omega_\rho\Delta=\Omega_\lambda\Delta\rightarrow 0$ as $\rho\rightarrow 0$.
Hence the density perturbations are given by
\be
\Delta^{\rm (DE)}=\Omega_\rho\Delta_0+ \Omega_\rho\Delta_1 a^{3\gamma -5}+
\sfrac{3\gamma}{3\gamma-7}\Omega_\rho\Sigma_0 a^2
\ee
and are suppressed as one approaches the dark energy solution.
\subsection{Vector Perturbations}
In the case of vector perturbations the propagation equations are
\bea
&&{W_*}'=(3\gamma-4){W_*}\,,\\
&&\Sigma '=-{\cal E}\,,\\
&&{\cal E}'={\cal E}-2\Sigma\,,\\
&&{\cal H}'={\bar{\Sigma}}'={\bar{\cal E}}'=0\,,\\
&&{\bar{Q}_*}'=8(3\gamma-4){W}_*\;,
\eea
which are subject to the following constraints
\bea
&&3\Sigma+6{Q_*}=4Z_*\;,\nonumber \\
&&3{\cal E}+6{Q_*}=2U_*+6(\Omega_\rho+2\Omega_\lambda)\Delta\;,\nonumber \\
&&\bar{\Sigma}={\cal H}\;, \nonumber \\
&&{\cal H}=8{W_*}-\bar{Q}_*=0\;,\nonumber \\
&&\bar{\cal H}=\bar{W_*}=0\;,\nonumber \\
&&\bar{\Sigma}=\bar{\cal E}=8{W_*}-\bar{Q_*}\;.
\eea
The inhomogeneity variables evolve according to
\bea
&&{\Delta}'=(3\gamma-3)\Delta-\gamma{Z_*}\;,\nonumber \\
&&{Z_*}'=6\sfrac{\gamma-1}{\gamma}\Delta -{U}_*\;,\nonumber \\
&&{Q_*}'=-Q_*-\sfrac{1}{3}{U_*}+4\sfrac{\gamma-1}{\gamma}\Delta\;,\nonumber \\
&&{U_*}'=-4Z_*+12\sfrac{\gamma-1}{\gamma}\Delta\;.
\eea
The solutions to this system are
\bea
&&\Sigma=\Sigma_0 a^2+\Sigma_1 a^{-1}\;,\nonumber \\
&&{\cal E}=-2\Sigma_0a^2+\Sigma_1a^{-1}\;,\nonumber \\
&&{\cal H}=\bar{\Sigma}=\bar{\cal E}={\cal H}_0\;,\nonumber \\
&&W=2W^*_0a^{3\gamma-3}\;,\nonumber \\
&&\bar{Q}_*=8W^*_0a^{3\gamma-4}+{\cal H}_0\;,
\eea
and
\bea
&&\Delta=\Delta_0+ \Delta_1 a^{3\gamma -5}
+\sfrac{3}{4}\sfrac{3\gamma}{3\gamma-7}\Sigma_0 a^2\;,\nonumber \\
&&Z_*=\sfrac{3\gamma-3}{\gamma}\Delta_0+ \sfrac{2}{\gamma}\Delta_1 a^{3\gamma -5}
+\sfrac{9}{4}\sfrac{3\gamma-5}{3\gamma-7}\Sigma_0 a^2\;,\nonumber \\
&&Q_*=\sfrac{2\gamma-2}{\gamma}\Delta_0+ \sfrac{4}{3\gamma}\Delta_1 a^{3\gamma -5}
+\sfrac{3\gamma-4}{3\gamma-7}\Sigma_0 a^2-\sfrac{1}{2}\Sigma_1 a^{-1}\;,\nonumber \\
&&U_*=\sfrac{6\gamma-6}{\gamma}\Delta_0+ \sfrac{4}{\gamma}\Delta_1 a^{3\gamma -5}
+\sfrac{9}{3\gamma-7}\Sigma_0 a^2\;,
\eea
where again $\gamma\neq\sfrac{7}{3}$.

This time there are six constants of integration:
$\Sigma_0,~\Sigma_1,~{\cal H}_0,~W^*_0,~Q^*_0,~\Delta_1$.

The vector modes of the physical quantities defined in (\ref{DE})
can then be found:
\bea
&&\Delta^{\rm (DE)}=\sfrac{\kappa^2\rho_0}{3H_0^2}\Delta_0 a^{4-3\gamma}
+\sfrac{\kappa^2\rho_0}{3H_0^2} \Delta_1 a^{-1}
+\sfrac{\kappa^2\rho_0}{4H_0^2} \sfrac{3\gamma}{3\gamma-7}\Sigma_0 a^{6-3\gamma}\;,\nonumber \\
&&Q^{\rm (DE)}=\sfrac{2\gamma-2}{3H_0\gamma}\Delta_0 a
+ \sfrac{4}{9H_0\gamma}\Delta_1 a^{3\gamma -4}
+\sfrac{4}{9H_0}\sfrac{3\gamma-4}{3\gamma-7}\Sigma_0 a^3\nonumber\\
&&~~~~~~~~-\sfrac{2}{9H_0}\Sigma_1\;,\nonumber \\
&&U^{\rm (DE)}=U_*\;.
\eea
\subsection{Tensor Perturbations}
The tensor parts of the propagation equations in the long
wavelength limit are:
\bea
&&\Sigma '=-{\cal E}\;,\nonumber \\
&&{\cal E}'={\cal E}-2\Sigma\;,\nonumber \\
&&{\cal H}'={\cal H}-\bar{\cal E}\;,\nonumber \\
&&{\bar{\Sigma}}'=\bar{\Sigma} -\bar{\cal E}\;,\nonumber \\
&&{\bar{\cal E}}'=2\bar{\cal E}-2\bar{\Sigma}\;,
\eea
subject to the following constraints:
\be
\bar{\cal H}=0\;, ~~ \bar{\Sigma}={\cal H}\;.
\ee
The solutions are
\bea
&&\Sigma=\Sigma_0 a^2+\Sigma_1 a^{-1}\;,\nonumber \\
&&{\cal E}=-2\Sigma_0a^2+\Sigma_1a^{-1}\;,\nonumber \\
&&{\cal H}=\bar{\Sigma}={\cal H}_0+{\cal H}_1 a^3\;,\nonumber \\
&&\bar{\cal E}={\cal H}_0-2{\cal H}_1 a^3\;,
\eea
with a constant of integration for each of the independent modes:
$\Sigma_0,~\Sigma_1,~{\cal H}_0,~{\cal H}_1$.
\section{High Energy Limit: The ${\cal F}_b$ Background}
The high energy limit is characterised by
$\sfrac{\rho^2}{\lambda}\gg\rho$ and $\sfrac{\rho^2}{\lambda}\gg\rho^*$ 
or $\Omega_\lambda \gg \Omega_\rho$ and $\Omega_\lambda \gg \Omega_{\rho^*}$ 
so this time we evaluate the perturbation equations in the limit
$(\Omega_\rho,\Omega_\lambda,\Omega_{\rho^*})\rightarrow(0,1,0)$.

This model corresponds to a stationary (equilibrium) point ${\cal F}_b$ in the 
phase space of homogeneous Bianchi models \cite{coley2,coley1}, as well as in the phase 
space of the special class of inhomogeneous $G_2$ cosmological models. In both 
cases ${\cal F}_b$ is found to be the source, or past attractor, for the generic 
dynamics for $\gamma>1$ ($\gamma=1$ is also included in the homogeneous case), 
consistent with~\cite{MS,SC2,SVF,SC1}. The stability of this result is now examined
through an analysis of the perturbation equations for this case.

The background scale factor $a$, Hubble function $H$ and deceleration
parameter $q$ of these models are given by
\be
a(t)=\left(t/t_0\right)^{1/{3\gamma}}\;,
~~H=H_0a^{-3\gamma},~q=3\gamma-1\;,
\ee
where again we fix an arbitrary initial
condition by choosing  $a_0=a(t_0)=1$.
The energy density behaves in the usual way:
\be
\rho=\rho_0a^{-3\gamma}\;.
\ee
From the Friedmann equation (\ref{f}) we find that
$H_0^2=\frac{\kappa^2}{6\lambda}\rho^2_0$.
\subsection{Scalar Perturbations}
The scalar propagation equations for this case reduce to
\bea
&&\Sigma '=(3\gamma-2)\Sigma -{\cal E}\;,\nonumber \\
&&{\cal E}'=(6\gamma-3){\cal E}-3\gamma\Sigma\;,
\eea
subject to the constraints
\bea
&&W=\bar{W}={\cal H}=\bar{\cal H}=\bar{\Sigma}=\bar{\cal E}=\bar{Q}=0\;,
\nonumber \\
&&2{Z_*}=2\Sigma +3{Q_*}\;,\nonumber \\
&&6\Delta = 2{\cal E}+3{Q_*}-{U_*}\;.
\eea
The scalar evolution equations for the inhomogeneity
variables are
\bea
&&{\Delta}'=(3\gamma-3)\Delta-\gamma{Z_*}\;,\nonumber \\
&&{Z_*}'=(3\gamma-2){Z}_*-3(3\gamma+1)\Delta -{U}_*\;,\nonumber \\
&&{Q_*}' =(3\gamma-3){Q_*}-\sfrac{1}{3}U_*-6\gamma\Delta\;,\nonumber \\
&&{U_*}'=(6\gamma-4){U_*}\;.
\eea
Solutions can again be easily obtained by solving the above
system of linear equations. They are
\bea
&&\Sigma=\Sigma_0 a^{6\gamma-2}+\Sigma_1 a^{3\gamma-3}\;,\nonumber \\
&&{\cal E}=-3\gamma \Sigma_0 a^{6\gamma-2}+\Sigma_1 a^{3\gamma-3}\;,
\eea
and
\bea
&&{\Delta}=\sfrac{1}{2}Q^*_0 a^{-3}-\sfrac{\gamma(3\gamma+1)}{6\gamma+1}\Sigma_0a^{6\gamma-2}
-\sfrac{\gamma}{2(6\gamma-1)} U^*_0a^{6\gamma-4}\;,\nonumber \\
&&{Z_*}=\sfrac{3}{2}Q^*_0 a^{-3}+\sfrac{(3\gamma+1)^2}{6\gamma+1}\Sigma_0a^{6\gamma-2}
+\sfrac{3\gamma-1}{2(6\gamma-1)} U^*_0a^{6\gamma-4}\;,\nonumber \\
&&{Q_*}=Q^*_0 a^{-3}+\sfrac{6{\gamma}^2}{6\gamma+1}\Sigma_0a^{6\gamma-2}
-\sfrac{2}{3}\Sigma_1 a^{3\gamma -3}\nonumber \\
&&~~~~~~~+\sfrac{3\gamma-1}{3(6\gamma-1)} U^*_0a^{6\gamma-4}\;,\nonumber \\
&&{U_*}=U^*_0 a^{6\gamma-4}
\eea
for $\gamma\neq\sfrac{1}{6}$, and
\bea
&&{\Delta}=(\sfrac{1}{2}Q^*_0 -\sfrac{1}{6}U^*_0)a^{-3}-\sfrac{1}{8}\Sigma_0a^{-1}
-\sfrac{1}{12} U^*_0 \ln a\,\, a^{-3}\,,\\&&
Z_*=\sfrac{3}{2}Q^*_0 a^{-3}+\sfrac{9}{8}\Sigma_0a^{-1}
-\sfrac{1}{4} U^*_0 \ln a\,\, a^{-3}\,,\\&&
{Q}_*=Q^*_0 a^{-3}+\sfrac{1}{12}\Sigma_0a^{-1}-\sfrac{2}{3}\Sigma_1 a^{-\sfrac{5}{2}}
-\sfrac{1}{6} U^*_0 \ln a\,\, a^{-3}\,,\\&&
{U}_*=U^*_0 a^{-3}
\eea
for $\gamma=\sfrac{1}{6}$.
The scalar parts of the physical quantities
$\Delta^{\rm (HE)},~ Q^{\rm (HE)},~U^{\rm (HE)}$
defined by (\ref{HE}) are given by
\bea
&&{\Delta}^{\rm (HE)}=\sfrac{\kappa^2\rho_0}{12H_0^2}Q^*_0 a^{3\gamma-3}
-\sfrac{\gamma(3\gamma+1)}{6\gamma+1}\sfrac{\kappa^2\rho_0}{6H_0^2}
\Sigma_0a^{9\gamma-2}\nonumber \\
&&~~~~~~~~~~~-\sfrac{\gamma}{6\gamma-1}\sfrac{\kappa^2\rho_0}{12H_0^2} U^*_0a^{9\gamma-4}\;,\nonumber \\
&&Q^{\rm (HE)}=\sfrac{1}{6H_0}Q^*_0 a^{3\gamma-4}
+\sfrac{1}{H_0}\sfrac{\gamma^2}{6\gamma+1}\Sigma_0 a^{9\gamma-3}
-\sfrac{1}{9H_0}\Sigma_1 a^{6\gamma -4} \nonumber \\
&&~~~~~~~~~~~+\sfrac{1}{18H_0}\sfrac{3\gamma-1}{6\gamma-1}U^*_0 a^{9\gamma-5}\;,\nonumber \\
&&U^{\rm (HE)}=\sfrac{1}{6}U^*_0 a^{6\gamma-4}
\eea
for $\gamma\neq\sfrac{1}{6}$, and
\bea
&&{\Delta}^{\rm (HE)}=\sfrac{\kappa^2\rho_0}{12H_0^2}(Q^*_0-\sfrac{1}{3}U^*_0)
a^{-\frac{5}{2}}-\sfrac{\kappa^2\rho_0}{48H_0^2}\Sigma_0a^{-\frac{1}{2}}\nonumber \\
&&~~~~~~~~~~-\sfrac{\kappa^2\rho_0}{72H_0^2} U^*_0\ln a \,\,a^{-\frac{5}{2}}\;,\nonumber \\
&&Q^{\rm (HE)}=\sfrac{1}{6H_0}Q^*_0 a^{-\frac{7}{2}}
+\sfrac{1}{6H_0}\sfrac{1}{12}\Sigma_0a^{-\frac{3}{2}}
-\sfrac{1}{9H_0}\Sigma_1 a^{-3}\;,\nonumber \\
&&~~~~~~~~~~~-\sfrac{1}{36H_0} U^*_0 \ln a\,\, a^{-\frac{7}{2}}\nonumber \\
&&U^{\rm (HE)}=\sfrac{1}{6}U^*_0 a^{-3}
\eea
for $\gamma=\sfrac{1}{6}$.

$\Sigma_0,~\Sigma_1,~Q^*_0,~U^*_0$ are arbitrary constants of integration corresponding
to the four independent modes.
\subsection{Vector Perturbations}
For vector perturbations the propagation equations are
\bea
&&{W_*}'=(3\gamma-4){W_*}\;, \nonumber \\
&&\Sigma '=(3\gamma-2)\Sigma -{\cal E}\;, \nonumber \\
&&{\cal E}'=(6\gamma-3){\cal E}-3\gamma\Sigma\;, \nonumber \\
&&{\cal H}'=(6\gamma-3){\cal H}-\bar{\cal E}\;, \nonumber \\
&&{\bar{\Sigma}}'=(6\gamma-3)\bar{\Sigma}-\bar{\cal E}\;, \nonumber \\
&&{\bar{\cal E}}'=(9\gamma-4)\bar{\cal E}-3\gamma\bar{\Sigma}\;, \nonumber \\
&&{\bar{Q}_*}'=(6\gamma-4)\bar{Q}_*-36\gamma^2{W}_*\;,
\eea
subject to the following constraints
\bea
&&4{Z_*}=3\Sigma+6{Q_*}\;,\nonumber \\
&&12\Delta=3{\cal E}+6{Q_*}-2U_*\;,\nonumber \\
&&{\cal H}=\bar{\Sigma}=\bar{\cal E}=12\gamma{W_*}-\bar{Q}_*\;,\nonumber \\
&&\bar{\cal H}=\bar{W_*}=0\;.
\eea
The propagation equations characterising the inhomogeneities are
\bea
&&{\Delta}'=(3\gamma-3)\Delta-\gamma{Z_*}\;,\nonumber \\
&&{Z_*}'=(3\gamma-2){Z}_*-3(3\gamma+1)\Delta -{U}_*\;,\nonumber \\
&&{Q_*}'=(3\gamma-3){Q_*}-\sfrac{1}{3}{U_*}-6\gamma\Delta\;, \nonumber \\
&&{U_*}'=(6\gamma-4){U_*}\;.
\eea
Solutions are easily obtained and are given by
\bea
&&\Sigma=\Sigma_0 a^{6\gamma-2}+\Sigma_1 a^{3\gamma-3}\;, \nonumber \\
&&{\cal E}=-3\gamma \Sigma_0 a^{6\gamma-2}+\Sigma_1 a^{3\gamma-3}\;,\nonumber \\
&&{\cal H}=\bar{\Sigma}=\bar{\cal E}={\cal H}_0 a^{6\gamma -4}\;,\nonumber \\
&&W=3\gamma W^*_0 a^{6\gamma-5}\nonumber\;,\\
&&{\bar{Q}}_*=12\gamma W^*_0 a^{3\gamma-4}-{\cal H}_0a^{6\gamma-4}\;,
\eea
and
\bea
&&{\Delta}^{\rm (HE)}=\sfrac{\kappa^2\rho_0}{12H_0^2}Q^*_0 a^{3\gamma-3}
-\sfrac{\gamma(3\gamma+1)}{6\gamma+1}\sfrac{\kappa^2\rho_0}{9H_0^2}
\Sigma_0a^{9\gamma-2}\nonumber \\
&&~~~~~~~~~~~-\sfrac{\gamma}{6\gamma-1}\sfrac{\kappa^2\rho_0}{12H_0^2} U^*_0a^{9\gamma-4}\;,\nonumber \\
&&{Z_*}=\sfrac{3}{2}Q^*_0 a^{-3}+\sfrac{3}{4}\sfrac{(3\gamma+1)^2}{6\gamma+1}\Sigma_0a^{6\gamma-2}
+\sfrac{3\gamma-1}{2(6\gamma-1)} U^*_0a^{6\gamma-4}\;,\nonumber \\
&&Q^{\rm (HE)}=\sfrac{1}{6H_0}Q^*_0 a^{3\gamma-4}
+\sfrac{1}{4H_0}\sfrac{3\gamma^2}{6\gamma+1}\Sigma_0a^{9\gamma-3}\nonumber \\
&&~~~~~~~~~~~-\sfrac{1}{12H_0}\Sigma_1 a^{6\gamma -4}
+\sfrac{1}{18 H_0}\sfrac{3\gamma-1}{6\gamma-1} U^*_0a^{9\gamma-5}\;,\nonumber \\
&&U^{\rm (HE)}=\sfrac{1}{6}U^*_0 a^{6\gamma-4}
\eea
for $\gamma\neq\sfrac{1}{6}$, and
\bea
&&{\Delta}^{\rm (HE)}=\sfrac{\kappa^2\rho_0}{12H_0^2}(Q^*_0-\sfrac{1}{3}U^*_0)
a^{-\frac{5}{2}}-\sfrac{\kappa^2\rho_0}{72H_0^2}\Sigma_0a^{-\frac{1}{2}}\nonumber \\
&&~~~~~~~~~~-\sfrac{\kappa^2\rho_0}{72H_0^2} U^*_0\ln a \,\,a^{-\frac{5}{2}}\;,\nonumber \\
&&Z_*=\sfrac{3}{2}Q^*_0 a^{-3}+\sfrac{27}{32}\Sigma_0a^{-1}
-\sfrac{1}{4} U^*_0 \ln a\,\, a^{-3}\;,\nonumber \\
&&Q^{\rm (HE)}=\sfrac{1}{6H_0}Q^*_0 a^{-\frac{7}{2}}
+\sfrac{1}{6H_0}\sfrac{1}{16}\Sigma_0a^{-\frac{3}{2}}
-\sfrac{1}{12H_0}\Sigma_1 a^{-3}\nonumber \\
&&~~~~~~~~~~~ -\sfrac{1}{36H_0} U^*_0 \ln a\,\, a^{-\frac{7}{2}}\;,\\
&&U^{\rm (HE)}=\sfrac{1}{6}U^*_0 a^{-3}
\eea
for $\gamma=\sfrac{1}{6}$. There are six constants of integration:
$\Sigma_0,~\Sigma_1,~{\cal H}_0,~W^*_0,~ Q^*_0,~U^*_0$.
\subsection{Tensor Perturbations}
The tensor parts of the propagation equations in the long
wavelength limit are:
\bea
&&\Sigma '=(3\gamma-2)\Sigma -{\cal E}\;, \nonumber \\
&&{\cal E}'=(6\gamma-3){\cal E}-3\gamma\Sigma\;, \nonumber \\
&&{\cal H}'=(6\gamma-3){\cal H}-\bar{\cal E}\;, \nonumber \\
&&{\bar{\Sigma}}'=(6\gamma-3)\bar{\Sigma}-\bar{\cal E}\;, \nonumber  \\
&&{\bar{\cal E}}'=(9\gamma-4)\bar{\cal E}-3\gamma\bar{\Sigma}\;,
\eea
subject to the constraints
\be
\bar{\cal H}=0,~~\bar{\Sigma}={\cal H}\,.
\ee
The solutions are given by
\bea
&&\Sigma=\Sigma_0 a^{6\gamma-2}+\Sigma_1a^{3\gamma-3}\;,\nonumber \\
&&{\cal E}=-3\gamma\Sigma_0a^{6\gamma-2}+\Sigma_1a^{3\gamma-3}\;,\nonumber \\
&&{\cal H}=\bar{\Sigma}={\cal H}_0a^{6\gamma-4}+{\cal H}_1a^{9\gamma-3}\;,\nonumber\\
&&{\bar{\cal E}}={\cal H}_0a^{6\gamma-4}-3\gamma{\cal H}_1a^{9\gamma-3}\;.
\eea
There are four constants of integration:
$\Sigma_0,~\Sigma_1,~{\cal H}_0,~{\cal H}_1$.
\begin{table*}
\caption{\label{tab:table2} Large\hs scale behaviour of the non\hs zero physically
relevant geometric and kinematic quantities for the different backgrounds in the limit
$a\rightarrow 0$. The values $S,V,T$ in brackets denote scalar, vector and tensor contributions.
One can easily see how the appropriately normalised perturbation quantities defined in
(\ref{LE})-(\ref{DE}) converge for wider ranges of $\gamma$ as $a\rightarrow 0$.}
\begin{tabular}{llccccccccccc}
Quantity  &mode& $0<\gamma<\sfrac{2}{9}$&$<\gamma<\sfrac{1}{3}$&$<\gamma<\sfrac{4}{9}$
&$<\gamma<\sfrac{5}{9}$& $<\gamma<\sfrac{2}{3}$
& $<\gamma<\sfrac{5}{6}$ & $<\gamma<1$ & $\gamma<\sfrac{10}{9}$
& $<\gamma<\sfrac{4}{3}$ & $<\gamma<\sfrac{5}{3}$ & $<\gamma<2$\\
\hline
\hline
\multicolumn{13}{c}{\textbf{Low energy limit}}\\ \hline
$\Sigma,~{\cal E}$&$\Sigma_0(S,V,T)$ & $\infty $& $\infty $&$\infty $& $\infty$
& $\infty$ & $0$ & $0$& $0$& $0$& $0$& $0$\\
&$\Sigma_1(S,V,T)$ &  $\infty $ & $\infty $&$\infty $&$\infty$ & $\infty $
&$\infty $& $\infty $& $\infty $& $\infty $& $\infty$ & $\infty $\\
\hline
${\cal H}$&${\cal H}_0(V,T)$& $\infty $&$\infty $& $\infty $ & $\infty$ & $\infty $
&$\infty $& $\infty $& $\infty $& $\infty $& $0$ & $0$\\
&${\cal H}_1(T)$& $\infty $&$\infty $&$\infty $& $\infty$  & $\infty$ & $0$ & $0$
& $0$& $0$& $0$& $0$\\
\hline
$W$&$W^*_0(V)$& $\infty$ & $\infty $&$\infty $&$\infty$ & $\infty$ & $\infty$ & $\infty$ & $\infty$
& $0$ & $0$ & $0$\\
\hline
$\Delta=\Delta^{\rm (LE)}$&$\Sigma_0(S,V)$ & $\infty $&$\infty $&$\infty $& $\infty$
& $\infty$ & $0$ & $0$& $0$& $0$& $0$& $0$\\
&$\Sigma_1,Q^*_0(S,V)$ &$\infty $&$\infty $&  $\infty $ & $\infty$ & $\infty $
&$\infty $& $\infty $& $\infty $& $\infty $& $\infty$ & $\infty $\\
&$U^*_0(S,V)$& $\infty $&$\infty $&$\infty$ & $\infty$ & $\infty$ & $\infty$
& $\infty$ & $\infty$& $\infty$ & $0$ & $0$\\
\hline
$Q^{\rm (LE)}$&$Q^*_0(S,V)$&$\infty $&$\infty $& $\infty$ & $\infty$ & $\infty$ &
$\infty$ & $\infty$ & $\infty$& $\infty$ & $0$ & $0$\\
&$U^*_0(S,V)$& $\infty$ & $\infty $&$\infty $&$\infty$ & $\infty$ & $\infty$ &
$\infty$ & $\infty$& $0$ & $0$ & $0$\\
\hline
$U^{\rm (LE)}$&$U^*_0(S,V)$&$\infty $&$\infty $& $\infty$ & $\infty$ & $\infty$ &
$\infty$ & $\infty$ & $\infty$& $\infty$ & $0$ & $0$\\
\hline
\hline
\multicolumn{13}{c}{\textbf{Dark energy limit}}\\ \hline
$\Sigma,~{\cal E}$&$\Sigma_0(S,V,T)$ & $0$& $0$ & $0$& $0$ & $0$ & $0$ & $0$& $0$&
$0$& $0$& $0$\\
&$\Sigma_1(S,V,T)$ &  $\infty $ &$\infty $&$\infty $& $\infty$ & $\infty $ &$\infty $
& $\infty $& $\infty $&$\infty $& $\infty $ & $\infty $\\
\hline
${\cal H}$&${\cal H}_0(V,T)$&  const & const & const & const & const &const& const&
const&const& const &const\\
&${\cal H}_1(T)$& $0$& $0$ & $0$& $0$ & $0$ & $0$ & $0$& $0$& $0$& $0$& $0$\\
\hline
$W$&$W^*_0(V)$& $\infty$ &$\infty $&$\infty $& $\infty$ & $\infty$ & $\infty$ &
$\infty$ & $0$& $0$ & $0$ & $0$\\
\hline
$\Delta^{\rm (DE)}$
\footnote{This quantity is suppressed by a factor of $\rho_0$, hence not significant
when
approaching the vacuum model $(R)$.}
&$\rho_0\Sigma_0(S,V)$ & $0 $& $0$ & $0 $& $0 $ & $0$ & $0$ & $0$& $0$& $0$& $0$& $0$\\
&$\rho_0\Delta_0(S,V)$& $0 $& $0$ & $0 $& $0 $ & $0$ & $0$ & $0$& $0$& $0$& $\infty $&
$\infty $\\
&$\rho_0\Delta_1(S,V)$&$\infty $ & $\infty$ & $\infty $ &$\infty $& $\infty $&
$\infty $&$\infty $& $\infty $ &$\infty $&$\infty $& $\infty $\\
\hline
$Q^{\rm (DE)}$&$\Sigma_0(S,V)$ & $0 $& $0$ & $0 $& $0 $ & $0$ & $0$ & $0$& $0$& $0$&
$0$& $0$\\
&$\Sigma_1(S,V)$ &  const &  const&  const& const & const &const& const& const&const&
const &const\\
&$\Delta_0(S,V)$& $0$& $0$ & $0 $& $0 $ & $0$ & $0$ & $0$& $0$& $0$& $0$& $0$\\
&$\Delta_1(S,V)$&$\infty $ &$\infty $&$\infty $& $\infty$ & $\infty $ &$\infty $&
$\infty $& $\infty $&$\infty $& $0$ & $0$\\
\hline
$U^{\rm (DE)}$&$\Sigma_0(S,V)$ & $0 $& $0$ & $0 $& $0 $ & $0$ & $0$ & $0$& $0$& $0$&
$0$& $0$\\
&$\Delta_0(S,V)$&  const&  const&  const & const & const &const& const& const&const&
const &const\\
&$\Delta_1(S,V)$&$\infty $ &$\infty $&$\infty $& $\infty$ & $\infty $ &$\infty $&
$\infty $& $\infty $&$\infty $& $\infty $ & $0$\\
\hline
\hline
\multicolumn{13}{c}{\textbf{High energy limit}}\\ \hline
$\Sigma,~{\cal E}$&$\Sigma_0(S,V,T)$ & $\infty $&$\infty $& $0$&$0$  & $0$ & $0$ &
$0$& $0$& $0$& $0$& $0$\\
&$\Sigma_1(S,V,T)$ &  $\infty $ & $\infty $&$\infty $&$\infty$ & $\infty $ &$\infty $
& $\infty $& $0$& $0$& $0$ & $0$\\
\hline
${\cal H}$&${\cal H}_0(V,T)$& $\infty $&$\infty $& $\infty $ & $\infty$ & $\infty $
&$0$& $0$& $0$& $0 $& $0$ & $0$\\
&${\cal H}_1(T)$& $\infty $& $\infty $&$0$&$0$  & $0$ & $0$ & $0$& $0$& $0$& $0$& $0$\\
\hline
$W$&$W^*_0(V)$& $\infty$ &$\infty $&$\infty$&$\infty$&$\infty$&$\infty$&$0$&$0$&$0$
&$0$ & $0$\\
\hline
$\Delta^{\rm (HE)}$&$\Sigma_0(S,V)$ & $\infty $&$0$& $0$&$0$  & $0$ & $0$ & $0$&
$0$& $0$& $0$& $0$\\
&$Q^*_0(S,V)$ & $\infty $&$\infty $& $\infty $&$\infty $  & $\infty $ & $\infty $ &
$\infty $& $0$& $0$& $0$& $0$\\
&$U^*_0(S,V)$& $\infty $&$\infty $& $\infty $&$0$  & $0$ & $0$ & $0$& $0$& $0$& $0$&
$0$\\
\hline
$Q^{\rm (HE)}$&$\Sigma_0(S,V)$ & $\infty $&$\infty $ &$0$&$0 $ & $0$ & $0$ & $0$&
$0$& $0$& $0$& $0$\\
&$\Sigma_1(S,V)$ &  $\infty $ &$\infty $&$\infty $& $\infty$ & $\infty $ &$0$& $0 $
& $0$& $0$& $0$ & $0$\\
&$Q^*_0(S,V)$ &  $\infty $ &$\infty $&$\infty $& $\infty$ & $\infty $ &$\infty $&
$\infty $& $\infty $
& $\infty $& $0$ & $0 $\\
&$U^*_0(S,V)$& $\infty$ &$\infty $&$\infty $& $\infty$ & $0$ & $0$ & $0$ & $0$& $0$
& $0$ & $0$\\
\hline
$U^{\rm (HE)}$&$U^*_0(S,V)$& $\infty$ &$\infty $&$\infty $& $\infty$ & $\infty$ &
$0$ & $0$ & $0$& $0$ & $0$ & $0$\\
\hline
\hline
\end{tabular}
\end{table*}
\section{Results and Discussion}
In the previous three sections we have developed and solved the perturbation
equations for the {\it low energy}, {\it dark energy} and 
{\it high energy} backgrounds, respectively.

The main results of our analysis is summarised in Table II, in which we present
the early time asymptotics $a\rightarrow 0$ of the physically relevant 
quantities for the different energy regimes. These physically relevant 
quantities are the harmonically decomposed components of the expansion 
normalised vorticity, shear, and the electric and magnetic parts of the 
Weyl tensor (\ref{defdimless1},\ref{defdimless2}), as well as the 
appropriate gradients of the energy density $\rho$, the non\hs local energy 
density $\rho^*$ and the non\hs local flux $q^*_a$ defined in 
(\ref{LE})-(\ref{HE}). The remaining quantities appearing in the previous 
sections are required to close the system of equations, but are otherwise 
of no particular physical importance.

We can see from Table II that in the low\hs energy regime the results 
from general relativity are recovered. In particular, we find the same 
decaying mode $\Sigma_1$ in both the shear $\Sigma$ and and density 
gradient $\Delta$, implying that in general relativity the flat RW models 
are {\it unstable} with respect to generic linear homogeneous and 
anisotropic perturbations into the past which was the problem outlined
in the introduction and partly the motivation for inflation.

In the dark energy limit, we find that for any value of $\gamma$ there 
is a quantity that diverges as $a\rightarrow 0$. The dark energy 
background is, however, an unstable equilibrium point in the state space 
of flat FRW models \cite{SC2}, and can therefore only be attained for 
very special initial conditions.

The main result of this analysis relates to the evolution of the 
perturbation quantities in the high energy background ${\cal F}_b$. 
We find that, unlike in GR, both shear and density gradient tend to 
zero at early times if $\gamma>4/3$. Thus the high\hs energy models 
isotropize into the past for realistic equations of state when we 
include generic linear inhomogeneous and anisotropic perturbations. 
\section{Conclusion}
In this paper we have given a comprehensive large\hs scale 
perturbative analysis of flat FRW braneworld models with 
vanishing cosmological constant, extending the work presented 
in \cite{dunsbyetal} by providing a complete analysis of 
{\it scalar}, {\it vector} and {\it tensor} perturbations for 
all the important stages of the braneworld evolution, namely 
the {\it low energy GR}, {\it high energy} and {\it dark energy} 
regimes. To make this precise we defined dimensionless variables 
that were specially tailored for each regime of interest. In this 
way we were able to clarify further the past asymptotic behaviour 
of these models and obtain results which exactly match the recent 
work of Coley {\it et. al.} \cite{CHL} on the spatially inhomogeneous 
$G_{2}$ braneworld models: isotropization towards the ${\cal F}_b$ 
model occurs for an equation of state parameter $\gamma > 4/3$.

\noindent {\bf Acknowledgements}
 PKSD and NG thank the NRF (South Africa) for financial support
and AC thanks NSERC (Canada) for financial support.

\end{document}